\documentclass[twocolumn,showpacs,preprintnumbers,amsmath,amssymb,nofootinbib]{revtex4-1}

\usepackage{graphicx}
\usepackage{amsmath}
\usepackage{amssymb}
\usepackage{epstopdf}
\usepackage{xcolor}
\usepackage{dcolumn}

\usepackage{threeparttable}
\usepackage{longtable}

\def\E51{\mathcal{E}}
\def\m10{\mathcal{M}}

\def\la{\lesssim}
\def\ga{\gtrsim}

\begin{document}

\title{PeV neutrinos from wind breakouts of type II supernovae}

\renewcommand{\thefootnote}{\fnsymbol{footnote}}

\author{Zhuo Li$^{1,2}$}
\affiliation{$^{1}$Department of Astronomy, School of Physics, Peking University, Beijing 100871, China;\\
$^{2}$Kavli Institute for Astronomy and Astrophysics, Peking University, Beijing 100871, China}

\footnotetext{Corresponding author(email: zhuo.li@pku.edu.cn)}

\date{\today}

\begin{abstract}
Recently, the rapid multiwavelength photometry and flash spectra of supernova (SN) 2013fs imply that the progenitor stars of regular type II SNe (SNe II) might be commonly surrounded with a confined dense stellar wind ejected by themselves with large mass loss rate few years before the SNe. Based on the assumption that the pre-SN progenitor stars of SNe II have a SN 2013fs-like wind, with mass loss rate $\dot{M}\sim3\times10^{-3}(v_w/100\rm km\,s^{-1})M_\odot\rm yr^{-1}$, we investigate the neutrino emission during the wind breakouts of SN shocks. We find that the regular SNe II can convert a fraction $\sim10^{-3}$ of their bulk kinetic energy into neutrino emission, which can contribute a significant fraction of the IceCube-detected neutrino flux at $\ga300$ TeV. Moreover, the $\la200$~TeV IceCube neutrinos can be accounted for by cosmic rays produced by the shocks of all SN remnants, losing energy in their host galaxies, i.e., starburst galaxies. The future follow-up observations of neutrinos by Gen2 and gamma-rays by CTA and LHAASO from nearby individual SNe II, within weeks after the explosions, will test this model.

\vspace{0.3cm}


\vspace{0.3cm}

\noindent
\end{abstract}

\maketitle

\section{Introduction}
Massive stars with initial mass larger than $8M_\odot$ end when their cores collapse. This triggers a supernova (SN), in which a strong shock is generated, propagates through the progenitor star, and ejects its envelope. The SN shock is radiation dominated inside the progenitor star, and a radiation flash is produced when the shock breakouts from the star \citep{2016arXiv160701293W}. Some observations of early SN radiation had been explained to be SN shock breakouts \citep[e.g.,][]{2006Natur.442.1008C,2008Natur.453..469S,2008Sci...321..223S,2010ApJ...724.1396O,2015ApJ...804...28G,2016ApJ...820...23G}. It is predicted that the shock becomes collisionless after breakouts, and can accelerate particles, which may interact with background nucleons leading to pion and hence neutrino and gamma-ray production \citep{2001PhRvL..87g1101W}.

If the progenitor star is surrounded by a dense stellar wind the SN shock will go through this dense medium rather than the interstellar medium, and the denser medium enhance $pp$ interaction rates and leads to more efficient pion and neutrino/gamma-ray production \citep{2011arXiv1106.1898K,2011PhRvD..84d3003M,2013ApJ...769L...6K,Zirakashvili:2015mua,2017MNRAS.470.1881P}. Whether there exists a dense stellar wind depends on the pre-SN evolution of massive stars, which is theoretically not well understood and difficult to observe. However, the properties of UV/optical emission and several X/gamma-ray flashes associated with SNe had suggested shock breakouts from dense stellar winds of progenitor stars \citep{2006Natur.442.1008C,2008Natur.453..469S,2010ApJ...724.1396O,2015ApJ...804...28G}. In particular, very recently it was reported that rapid follow-up photometry and spectroscopy observations of SN 2013fs map the immediate environment of the progenitor star and establish that it was surrounded by a confined, dense circumstellar material \citep{2017NatPh..13..510Y}. The observations indicate that SN 2013fs is a regular type II SN (SN II), thus it may be common that red supergiant stars (RSGs), progenitors of SNe II, ejected a dense wind at a high rate just $\sim$ yr before the SN explosions \citep{2017NatPh..13..510Y}.

Diffuse TeV-PeV neutrinos had been first detected by IceCube \citep{2013PhRvL.111b1103A,ic13}, but the origin is still unknown. IceCube did not find point sources yet for 7 yr search \citep{2017ApJ...835..151A}, and the Galactic, blazar and gamma-ray burst origins of the bulk diffuse neutrinos had been strongly disfavored \citep{2014JCAP...11..028W,2016SCPMA..59a5759W,2017JCAP...03..024Z,2017ApJ...835...45A,2017ApJ...843..112A,Murase:2016gly}. The latest IceCube result from the south hemisphere \citep{2017arXiv171001191I} hints that the neutrino spectrum may not be a featureless single power law, but consist of more spectral components: the spectrum beyond a few $100$'s~TeV is flat $E_\nu^2\phi_\nu\propto E_\nu^0$, consistent with the north hemisphere muon track events \citep{2016ApJ...833....3A,2017arXiv171001191I}, but below $100$~TeV the flux is enhanced abruptly by a factor of $\sim4$ \citep[][see Fig. \ref{fig:diffuse}]{2017arXiv171001191I}.

Motivated by these findings, we investigate the TeV-PeV neutrino production from normal SNe II, assuming that the pre-SN progenitor stars are commonly surrounded by dense winds, up to at least $\sim10^{15}$cm, ejected by themselves. We find that it is likely that the wind breakouts of SN shocks can account for at least a significant fraction of the IceCube detected diffuse neutrinos at $\ga300$~TeV, and the neutrinos below $\sim200$~TeV can be contributed by SN remnant (SNR) shock-produced cosmic rays (CRs) interacting with interstellar medium in the host galaxy. Note, Ref. \cite{Murase:2017pfe} recently had calculated the neutrino flux from a SN II with SN 2013fs-like wind, but focused on the next SN in the Milky Way.

\section{Dynamics}
Consider that the pre-SN progenitor star is surrounded by a stellar wind with the density $\rho=\dot{M}/4\pi R^2v_w\equiv AR^{-2}$, where $\dot{M}$ is the wind mass loss rate, $v_w$ is the wind velocity, and $R$ is the radius. According to the measurement of SN 2013fs \citep{2017NatPh..13..510Y}, we take $A=1.5\times10^{15}A_\star\rm g\,cm^{-1}$ for $\dot{M}=3\times10^{-3}M_\odot\rm yr^{-1}$ and $v_w=100~\rm km\,s^{-1}$, and the wind is confined but extends up to a distance of $R_w\approx10^{15}$cm.

The SN explosion ejects the progenitor's stellar envelope. Typically for a SN II, the total ejecta mass is $M=10\m10 M_\odot$, and the bulk kinetic energy is $E_k=10^{51}\E51$ erg, thus we take them as the normalization \citep[as][]{1999ApJ...510..379M}, hence the bulk velocity is $v_b=\sqrt{2E_k/M}=3.2\times10^8\E51^{1/2}\m10^{-1/2}\rm cm\,s^{-1}$. When the SN shock propagates down the density gradient of the outer part of the stellar envelope, it accelerates and the swept-up material is also accelerated and ejected. After the shock breaks through the stellar surface, the SN ejecta left behind has more energy for slower shell. The kinetic energy of ejecta with velocity larger than $v$ is given by
\begin{equation}\label{eq:ejecta_velocity}
  E_{ej}(>v)= E_k(v/v_b)^{-\chi}~~(v\geq v_b)
\end{equation}
where $\chi=3+(5/n)$, with $n=3/2$ and 3 for convective and radiative envelopes, respectively \citep{1999ApJ...510..379M}. We take $\chi=6$ for RSG in the following, but we also try $\chi=5$ (for blue supergiant stars; BSGs) which gives negligible change of the results.

After acceleration in the steep gradient of the stellar outer envelope, the SN shock will be decelerated in the wind. When the shock propagates to a radius $R$ in the wind, the total energy of the shock-swept-up wind material is $E_s=u^2\int_0^R 4\pi r^2\rho dr=4\pi ARu^2$ with $u$ the postshock fluid velocity. This shock energy is provided by the SN ejecta with velocity $v\geq u$, thus let $E_s(u)=E_{ej}(>v)|_{v=u}$, resulting in the dynamical evolution of the SN shock in the wind,
\begin{equation}\label{eq:dyn}
  u=\left(\frac{E_kv_b^6}{4\pi A}\right)^{1/8}R^{-1/8}.
\end{equation}
Note this description is available for early stage when the SN shock propagates in the wind where $u>v_b$. The dynamical evolution with $u<v_b$ should be derived by $E_s=E_k$, available at large radii but not relevant here.

For a strong shock, the shock velocity is $v_s\simeq u$. As the SN shock propagates to the point where the radiation diffusive velocity becomes larger than the shock velocity, the radiation escapes from the shock and produces a shock breakout flash. This happens when the optical depth of the material ahead of the shock is $\tau_{br}=c/v_s$ \citep{1999ApJ...510..379M}. If the wind is dense enough the shock breakout happens in the wind. The optical depth of wind at radius $R$ is given by $\tau_w=(\rho/m_p)\sigma_TR$. Equating $\tau_w=\tau_{br}$ gives the breakout radius and velocity relation $R_{br}=(A\sigma_T/m_p)v_{br}/c$, which, combined with eq (\ref{eq:dyn}), further gives $ v_{br}=\left(\frac{E_kv_b^6m_pc}{4\pi \sigma_TA^2}\right)^{1/9}
=1.1\times10^9\frac{\E51^{4/9}}{A_\star^{2/9}\m10^{1/3}}\rm cm\,s^{-1}$
and then
$  R_{br}=2.2\times10^{13}\frac{A_\star^{7/9}\E51^{4/9}}{\m10^{1/3}}\rm cm$.
If the wind optical depth is smaller than $\tau_{br}$, the shock breakout occurs in the stellar surface.

\section{Particle acceleration and energy loss}
Initially the SN shock is radiation-mediated inside the stellar envelope. Once the radiation escapes, or even before that \citep{2011arXiv1106.1898K,2015MNRAS.449.3693G}, the shock can no longer be mediated by radiation. Since ion plasma frequency is many orders of magnitude larger than the other relevant frequencies, the shock is expected to become collisionless, and be mediated by collective plasma instabilities \citep{2001PhRvL..87g1101W}. The collisionless shock starts to accelerate particles via diffusive shock acceleration (DSA) \citep{1987PhR...154....1B}.

Normalizing the diffusion coefficient to the Bohm value, the acceleration timescale of protons with energy $E_p$ can be given by $t_{acc}=f_BE_pc/v_s^2eB$, where $f_B\ga1$ is a constant accounting for the uncertainty of particle diffusion, and $B=\sqrt{8\pi \epsilon_B\rho v_s^2}$ is the postshock magnetic field strength, with $\epsilon_B$ being the fraction of energy carried by magnetic field. We take $\epsilon_B=10^{-2}\epsilon_{B,-2}$ as the typical value, as estimated by X-ray filaments in young SNRs \citep{Volk:2004vi,Uchiyama:2007zz}. As for the uncertainty of diffusion, some X-ray observations of young SNRs already indicate fast acceleration close to Bohm limit \citep{Uchiyama:2007zz,WangLi}, i.e., $f_B\sim$ few.

In the case of dense stellar winds considered here, the high energy protons will lose energy mainly by $pp$ interactions with background medium, producing pions. The $pp$ energy loss timescale is $t_{pp}=[0.5(\rho_s/m_p)\sigma_{pp}c]^{-1}$, where $\rho_s=4\rho$ is the postshock density, and the $pp$ pion production cross section is $\sigma_{pp}(E_p)=3.43\times10^{-26}\theta(E_p)$cm$^2$, with $  \theta(E_p)=1+0.055l+0.0073l^2$
and $l=\ln(E_p/1~\rm TeV)$ \citep{2006PhRvD..74c4018K}. The $pp$ energy loss timescale $t_{pp}\propto R^2/\theta(E_p)$ weakly depends on $E_p$ due to $\theta(E_p)$, and increases with $R$ faster than the dynamical timescale $R/v_s\propto R^{9/8}$.

The particle acceleration suffers from both limited shock expansion time and the energy loss. If the maximum energy of accelerated protons is limited by $pp$ energy loss, by equating $t_{acc}=t_{pp}$, we have the maximum energy
\begin{equation}\label{eq:max_pp}
  E_{p,\max}^{pp}(R)=35\frac{\E51^{3/2}\epsilon_{B,-2}^{1/2}}{A_\star^{7/8}\m10^{9/8}\theta f_B}R_{15}^{5/8}\rm PeV
\end{equation}
where $R=10^{15}R_{15}$cm. If limited by the dynamical time of the shock, then $t_{acc}=R/v_s$ gives
\begin{equation}\label{eq:max_dyn}
  E_{p,\max}^{dyn}(R)=94\frac{A_\star^{1/4}\E51\epsilon_{B,-2}^{1/2}}{\m10^{3/4}f_B}R_{15}^{-1/4}\rm PeV.
\end{equation}
At radius $R$, the maximum proton energy should be $E_{p,\max}(R)=\min(E_{p,\max}^{pp},E_{p,\max}^{dyn})$. Since $E_{p,\max}^{pp}$ increases with $R$ but $E_{p,\max}^{dyn}$ decreases, $pp$ energy loss is more important constraint at small radii, but the dynamical time limit more important at large radii. A highest value of $E_{p,\max}(R)$ appears if $E_{p,\max}^{pp}=E_{p,\max}^{dyn}$ (i.e., $t_{acc}=t_{pp}=R/v_s$), which reads
\begin{equation}\label{eq:Emax}
  E_{p,\max}\theta^{2/7}(E_{p,\max})=71\frac{\E51^{8/7}\epsilon_{B,-2}^{1/2}}{A_\star^{1/14}\m10^{6/7}f_B}\rm PeV,
\end{equation}
insensitive of $A$ and $\epsilon_B$. Note, this is available if the wind is not confined but extends to large distance. However, if the wind is confined to $R_w\approx10^{15}$cm, the highest proton energy is determined by eq (\ref{eq:max_pp}), $E_{p,\max}\theta \sim35f_B^{-1}$~PeV. So, the proton energy could reach $\la100$ PeV, for fast particle acceleration, $f_B\sim$ few.

Since the energy loss rate ($t_{pp}^{-1}$) increases with the proton energy, for a certain $R$ there should be a critical proton energy $E_{p,\rm loss}(R)$, above which protons significantly lose energy by $pp$ interactions. This can be defined by requiring $t_{pp}(E_{p,\rm loss},R)=R/v_s(R)$. Since the work concerns mainly about efficient neutrino production, we are more interested in small radii. At small radii where the maximum proton energy is constrained by $pp$ energy loss rather than the dynamical time, we have $t_{pp}(E_{p,\max}^{pp})<t_{dyn}$. Since $t_{pp}(E_{p,\rm loss})=t_{dyn}$ as defined, we have $t_{pp}(E_{p,\max}^{pp})<t_{pp}(E_{p,\rm loss})$, and hence $E_{p,\rm loss}(R)<E_{p,\max}^{pp}(R)$.

On the other hand, equivalently, we can define by $t_{pp}=R/v_s$ a critical radius for a certain $E_p$,
\begin{equation}\label{eq:pploss_radius}
  R_{pp}(E_p)=3.1\times10^{15}\frac{A_\star^{9/7}\m10^{3/7}\theta^{8/7}}{\E51^{4/7}}\rm cm,
\end{equation}
within which the proton can lose energy efficiently by $pp$ interactions. $R_{pp}$ weakly increases with $E_p$ due to function $\theta(E_p)$, so higher energy protons can efficiently lose energy at somewhat larger radii of lower wind density. Since $\theta>1$, $R_{pp}(E_p)>3.1\times10^{15}$cm; at $R<3.1\times10^{15}$cm, all accelerated protons lose energy efficiently by $pp$ interactions.


For protons to be accelerated and lose energy efficiently by $pp$ interactions, it is required that $R_{pp}$ is larger than the shock breakout radius, which is $R_{br}$ if breakout from the wind, or about the stellar radius $R_\star$ if breakout from the stellar surface, thus $R_{pp}(E_p)>\max(R_{br},R_\star)$, which reads
$
  A_\star>\max\left(6\times10^{-5}\frac{\E51^{2}}{\m10^{3/2}\theta^{9/4}}, 0.03\frac{\E51^{4/9}R_{\star,500}^{7/9}}{\m10^{1/3}\theta^{8/9}}\right)
$,
where we take $R_\star=500 R_{\star,500}R_\odot$ as typical value for RSGs. This requirement for $A$ is easily satisfied in the case of the dense wind in SN 2013fs.

\section{Neutrino production}
\subsection{Individual SNe}
It is assumed that in a strong shock the particles are accelerated to follow a flat energy distribution, $dN_p/dE_p\propto E_p^{-s}$ ($E_{p,\min}<E_p<E_{p,\max}$) with $s\approx2$, which is theoretically predicted and consistent with observed nonthermal emission from SNR shocks \citep{1987PhR...154....1B}. The accelerated particles can carry a fraction $\xi\ga$ tens percents of the shock energy for efficient particle acceleration in the latest DSA theories, whereas $\xi\sim0.1$ is required for explanation of the origin of Galactic CRs by SNRs \citep[e.g.,][]{Ptuskin:2010zn,Zirakashvili:2015mua}. We here conservatively take $\xi=10^{-1}\xi_{-1}$. When the shock propagates to radius $R$ with velocity $v_s(R)$, the shock energy is $E_{ej}(>v)|_{v=v_s(R)}$, thus the energy distribution of all accelerated protons is given by
\begin{equation}\label{eq:cr_energy}
    E_p^2\frac{dN_p}{dE_p}=\frac{\xi E_{ej}(>v)|_{v=v_s(R)}}{\ln(E_{p,\max}(R)/E_{p,\min})}.
\end{equation}
Since $E_{p,\max}(R)$ and $E_{p,\min}$ do not change significantly with $R$, we simply take $1/\ln(E_{p,\max}(R)/E_{p,\min})\sim1/7\ln(10)$. 
Protons with energy in the range of $E_{p, \rm loss}(R)<E_p<E_{p, \max}(R)$ essentially lose all their energy by pion production, then the charged pion decays lead to neutrino production. We assume the neutrino flavor ratio after mixing in propagation is $\nu_e:\nu_\mu:\nu_\tau\approx1:1:1$.
The spectrum of produced neutrinos (per flavor) is
\begin{equation}\label{eq:cr2nu}
  E_\nu^2\frac{dN_\nu}{dE_\nu}=\frac16 E_p^2\frac{dN_p}{dE_p}
\end{equation}
where the factor $1/6$ results from the facts that a fraction of 2/3 of the proton energy goes to charged pions in $pp$ interactions (i.e., $\pi^+:\pi^-:\pi^0\approx1:1:1$), and that each neutrino carries a fraction 1/4 of the charged pion energy. Moreover, in each $pp$ interaction the produced charged pion energy is approximately a factor $1/5$ of the primary proton energy, thus the produced neutrino energy is about $E_\nu\approx E_p/20$.

We consider two cases of the stellar wind: one is extended wind (EW) to radius $R_{pp}(E_{p,\max})\sim6\times10^{15}$cm, and another is confined wind (CW) only up to radius $R_w\approx10^{15}R_{w,15}$cm. In the EW case, during the whole wind breakout event, the neutrino emission at $E_\nu$ is dominated by protons of $E_p=20E_\nu$ that are accelerated at radius $R=R_{pp}(E_p=20E_\nu)$. This is because at larger radii where the shock velocity is smaller the shock obtains larger energy from the slower SN ejecta shell. But the radius is limited to $R\la R_{pp}(E_p)$, since at even larger radii the proton energy loss is negligible. Thus, substituting $R=R_{pp}(E_p)$ (eq.\ref{eq:pploss_radius}) into eqs. (\ref{eq:dyn}) and then (\ref{eq:ejecta_velocity}), one obtains the fraction of shock energy in the bulk ejecta energy,
$\eta^{\rm EW}\equiv E_{ej}(>v)|_{v=v_s(R_{pp})}/E_k=[v_s(R_{pp})/v_b]^{-6}$, i.e.,
\begin{equation}
  \eta^{\rm EW}=2.1\times10^{-2}\frac{A_\star^{12/7}\theta^{6/7}}{\E51^{3/7}\m10^{3/7}}.
\end{equation}
In the CW case, all accelerated protons significantly lose energy within $R_w$, thus $\eta^{\rm CW}\equiv E_{ej}(>v)|_{v=v_s(R_w)}/E_k=[v_s(R_w)/v_b]^{-6}$, i.e.,
\begin{equation}
  \eta^{\rm CW}=0.91\times10^{-2}\frac{A_\star^{3/4}R_{w,15}^{3/4}}{\m10^{3/4}}.
\end{equation}

\begin{figure}[t]
  \centering
  \includegraphics[width=\columnwidth]{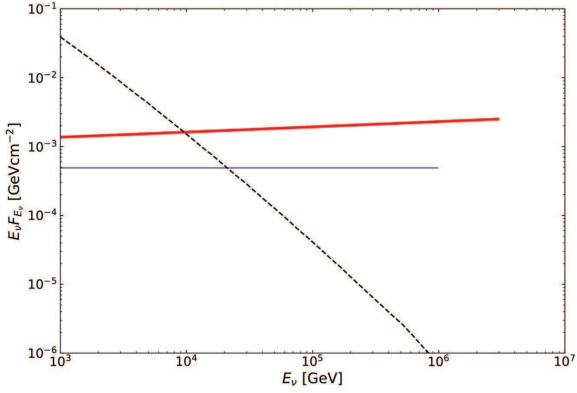}\\
  \caption{The neutrino fluence (per flavor) as function of neutrino energy for a SN II at $d_L=10$~Mpc. The thin (blue) and thick (red) solid lines are the confined-wind (CW) and extended-wind (EW) cases, respectively. The microphysical parameters assumed are $\xi_{-1}=\epsilon_{B,-2}=f_B=1$. The dashed line is the atmospheric $\mu$ neutrino background averaged over zenith angles within 1 degree \citep{2011PhRvD..83a2001A} and integrated over 50 days.}\label{fig:single}
\end{figure}

For EW case, plugging $\eta E_k$ into eq. (\ref{eq:cr_energy}) and with help of eq. (\ref{eq:cr2nu}), the emitted neutrino spectrum (time-integrated) can be given by\footnote{The result is comparable to that by \cite{Murase:2017pfe} for single SN II-P, although the details of two models are different, e.g., different shock dynamics adopted.}
\begin{equation}\label{eq:nuflux1}
  E_\nu^2\frac{dN_\nu}{dE_\nu}=2.2\times10^{46} \frac{\xi_{-1}A_\star^{12/7}\E51^{4/7}\theta^{6/7}}{\m10^{3/7}}\rm erg.~~~~(EW)
\end{equation}
This spectrum weakly depends on $E_\nu$ through $\theta(20E_\nu)$, and extends to a cutoff neutrino energy, corresponding to the maximum proton energy, $E_{\nu,\max}=E_{p,\max}/20$. By eq.(\ref{eq:Emax}) we have $E_{\nu,\max}\theta^{2/7}(20E_{\nu,\max})\approx3.5$~PeV. For CW case similarly we obtain
\begin{equation}\label{eq:nuflux2}
  E_\nu^2\frac{dN_\nu}{dE_\nu}=9.4\times10^{45} \frac{\xi_{-1}A_\star^{3/4}\E51 R_{w,15}^{3/4}}{\m10^{3/4}}\rm erg,~~~~(CW)
\end{equation}
independent of $E_\nu$, and $E_{\nu,\max}\theta(20E_{\nu,\max})\approx1.8$~PeV (eq.\ref{eq:max_pp}). Fig \ref{fig:single} shows, for a SN with luminosity distance of $d_L=10$~Mpc, the spectrum of the neutrino fluence, i.e. the flux integrated over the whole duration, $E_\nu F_{E_\nu}=(E_\nu^2dN_\nu/dE_\nu)/4\pi d_L^2$.

The duration of neutrino emission can be estimated by $ T\simeq\int_0^{R_{pp}}dR/v_s=(8/9) R_{pp}/v_s(R_{pp})$ for EW case, i.e.,
\begin{equation}
  T\simeq52\frac{A_\star^{11/7}\m10^{6/7}\theta^{9/7}}{\E51^{8/7}}\rm day.~~~~(EW)
\end{equation}
Replacing $R_{pp}$ with $R_w$ we have for CW case,
\begin{equation}
  T\simeq46\frac{A_\star^{79/56}\m10^{45/56}R_{w,15}^{1/8}\theta^{8/7}}{\E51^{15/14}}\rm day.~~~~(CW)
\end{equation}

\subsection{Diffuse emission}
\begin{figure}
  \centering
  \includegraphics[width=\columnwidth]{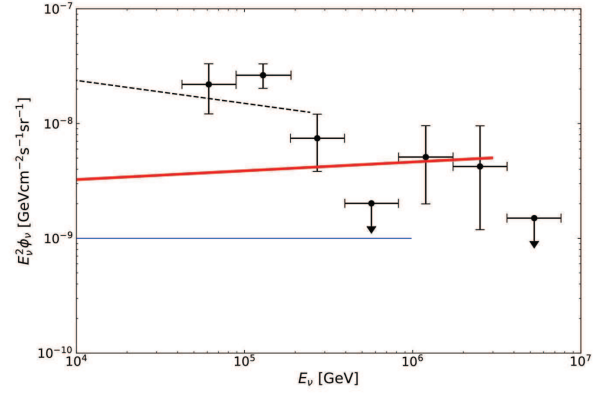}\\
  \caption{The diffuse neutrino intensity (per flavor) as function of neutrino energy. The solid lines are the contribution from wind breakouts of SNe II for confined-wind (CW; thin and blue) and extended-wind (EW; thick and red) cases, respectively. The dashed line is the contribution from SNR shock-produced CRs propagating and efficiently losing energy in host galaxies. The data show the latest results of IceCube's high energy starting events \citep{2017arXiv171001191I}.}\label{fig:diffuse}
\end{figure}

The diffuse neutrino intensity from all SNe II in the universe can be calculated by integration over the SN rate history, and given by
$
E_\nu^2\phi_\nu=\frac c{4\pi}\zeta\dot{\rho}t_H E_\nu^2\frac{dN_\nu}{dE_\nu}
$
where $t_H$ is the Hubble timescale, $\dot{\rho}$ is the volumetric rate of local SNe, and $\zeta$ accounts for the effect of SN rate density evolution with redshift $z$. The core-collapse SN rate should follow the star formation rate (SFR), in which case the factor is calculated to be about $\zeta\simeq3$ \citep{wbbound}.
The volumetric rates of nearby core-collapse SNe had been measured, $\dot{\rho}=0.7\times10^{-4}\rm Mpc^{-3}yr^{-1}$\citep{2011MNRAS.412.1473L}, most of which are SNe II. Using $t_{H}=10$ Gyr, and with help of eqs. (\ref{eq:nuflux1}) or (\ref{eq:nuflux2}), the diffuse neutrino intensity is given by
\begin{equation}
  E_\nu^2\phi_\nu=2.3\times10^{-9}\frac{\xi_{-1}A_\star^{12/7}\E51^{4/7}\theta^{6/7}}{\m10^{3/7}}\rm GeV\,cm^{-2}s^{-1}sr^{-1}~~~~(EW)
\end{equation}
or
\begin{equation}
  E_\nu^2\phi_\nu=1.0\times10^{-9}\frac{\xi_{-1}A_\star^{3/4}\E51 R_{w,15}^{3/4}}{\m10^{3/4}}\rm GeV\,cm^{-2}s^{-1}sr^{-1},~~~~(CW)
\end{equation}
extending up to the neutrino maximum energy, $E_{\nu,\max}$. In CW case the neutrino flux at PeV is about a fraction $\sim1/3$ of the IceCube detected one, whereas the EW case can well match the IceCube data at $E_\nu\ga300$~TeV, as shown in Fig \ref{fig:diffuse}.

We take a flat CR spectrum in the SN shocks, which is predicted by DSA theory for strong shocks in the test particle assumption, but the theory is with uncertainty. A harder (softer) CR spectrum will enhance (reduce) the neutrino flux at the high energy end. For example, for a spectral index of $s>2$, the neutrino flux is lower by a factor of $(20E_\nu/E_{p,\min})^{p-2}/(p-2)\ln(E_{p,\max}/E_{p,\min})$ compared with the case of $s=2$. For $s\approx2.2$, this factor is $\sim5$ for neutrino flux at $E_\nu\ga300$~TeV.


\section{Low energy neutrinos from CR propagation}
The relatively larger neutrino flux at $E_\nu\la100$~TeV from the latest IceCube data seems difficult to be accounted for by wind breakouts of SNe II. Here we show that the low energy flux can be naturally explained by the contribution of the SN shocks after wind breakouts. The shock eventually will go through the wind and be driven into the wind bubble or the interstellar medium (ISM). In the late time the SNR shock continues accelerating particles, which, after escape from the shock, interact with the ISM during propagation in the host galaxy. The star formation is dominated by starburst galaxies (SBGs) in the whole star formation history, thus most core-collapse SNe also occur in SBGs, where the relatively larger density and stronger magnetic field in the ISM, compared to normal galaxies, make them strong candidates of neutrino producers \citep{loebwaxman}.

Here we follow Refs. \citep{katz13,2014JCAP...11..028W} to estimate the low energy neutrino flux. Fermi-LAT had detected GeV gamma-rays from several nearby SBGs, and showed that the ratio of gamma-ray luminosity to SFR for individual SBGs is constant. Applying this constant to all SBGs in the universe, and using the measured SFR evolution with redshift, we can calculate the total GeV gamma-ray flux from all SBGs in the universe. By the correlation between gamma-ray and neutrino production in $pp$ interactions, we can derive the diffuse neutrino intensity at $E_\nu=0.5$~GeV, $E_\nu^2\phi_\nu=1.7\times10^{-7}(\zeta/3)\rm GeV\,cm^{-2}s^{-1}sr^{-1}$ \citep{2014JCAP...11..028W}. Extrapolating this flux with a power law to higher energy, we have
\begin{equation}\label{eq:lowE_diffuse}
  E_\nu^2\phi_\nu\approx1.5\times10^{-8}\frac{\zeta}{3}(E_\nu/{\rm 100 TeV})^{-0.2}\rm
  GeV\,cm^{-2}s^{-1}sr^{-1}.
\end{equation}
Here a neutrino spectrum of $dN_\nu/dE_\nu\propto E_\nu^{-2.2}$ is assumed,
consistent with measured GeV-TeV gamma-ray spectra of the nearby SBGs \citep{fermi-sfg,2016ApJ...821L..20P}.

The maximum proton energy produced by SNR shock corresponds to the deceleration radius where the shock swept-up medium mass is comparable to the ejecta mass and the shock starts to decelerate significantly. With the SNR shock dynamics, $E_k\simeq (4/3)\pi R^3nm_pv_s^2$, and equating the acceleration time and dynamical time at the deceleration radius, one obtains a limit to the energy of protons produced by the SNR shock during its whole evolution,
$
  E_p\la5\frac{\E51\epsilon_{B,-2}^{1/2}n_{-1}^{1/6}}{\m10^{2/3}f_B}\rm PeV,
$
where $n=10^{-1}n_{-1}\rm cm^{-3}$ is the medium density for the SNR shock. This limit leads to a spectral cutoff at $E_\nu\simeq250$~TeV in the neutrino spectrum (eq.\ref{eq:lowE_diffuse}). As shown in Fig \ref{fig:diffuse}, the contribution by SNR shock-produced CRs can reasonably account for the IceCube data at $E_\nu\la300$~TeV.

Recently there seems to be a tension between the SBG model of neutrino origin and the Fermi-measured extragalactic gamma-ray background (EGB), because the accompanying $pp$-induced gamma-ray emission should satisfy the EGB measurement. Our model predicts a neutrino flux of $\sim10^{-8}\rm GeV\,cm^{-2}s^{-1}sr^{-1}$ with a flat spectrum, $dN_\nu/dE_\nu\propto E_\nu^{-2.2}$ (see Eq.\ref{eq:lowE_diffuse}), which is indeed consistent with the Fermi-measured isotropic gamma-ray background (IGB), i.e. the EGB with resolved point sources subtracted \citep{Murase:2013rfa}. See, however, e.g., \cite{Bechtol:2015uqb}, a tension may rise if the non-blazar originated EGB is constrained to be less than a half of the IGB in the range of 50~GeV$-1$~TeV.

\section{Conclusion and discussion}
According to the recent results that the progenitor stars of SNe II may be commonly surrounded with dense winds ejected by themselves, we investigate the neutrino emission when the SN shocks breakout from winds. We find that the wind breakouts of SN II shocks can convert a fraction $\eta\xi\sim10^{-3}\xi_{-1}$ of the bulk kinetic energy into neutrinos, and can account for a significant fraction, $\sim1/3$, of the IceCube neutrinos at $\ga300$~TeV, if assuming a SN 2013fs-like, confined wind. If the wind extends to $R>R_{pp}(E_{p,\max})\sim6\times10^{15}$cm (EW case), or the CR acceleration in the SN shock is more efficient, $\xi\sim0.3$, the neutrino flux and spectrum can well fit the IceCube data at $\ga300$~TeV. Furthermore, the IceCube neutrinos below few hundreds TeV can be explained by SNR shock-produced CRs in SBGs. In this picture, the high energy neutrinos above a few hundreds TeV are contributed by transients of $\sim50$days, whereas the low energy neutrinos below few hundreds TeV are produced in a more steady process.

A subset of SNe II, SNe IIn, have even denser and more extended circumstellar material. They are expected to convert a larger fraction of CR energy into neutrinos, so although they are a small subset, their contribution to the diffuse neutrino flux could be $\sim10^{-9}\rm GeV\,cm^{-2}s^{-1}sr^{-1}$ \citep{2011PhRvD..84d3003M}, comparable to the regular SNe II. Thus the total contribution from both regular SNe II (CW case) and SNe IIn may account for the IceCube diffuse neutrino flux at $\ga300$~TeV.

One may worry whether the model satisfies the observational limits on neutrino doublets by IceCube. No neutrino doublet detected in the 4-yr IceCube data sets limits on the source luminosity and density, i.e., $E_\nu L_{E_\nu}\la10^{42}\rm erg\,s^{-1}$ and $n_0\ga10^{-7}\rm Mpc^{-3}$, respectively \citep{Murase:2016gly}. In our model, the $\ga300$~TeV neutrinos are from transient SNe II. During the observational period $\tau=4$~yrs, the number of SNe II that explode is $\dot{\rho}\tau\sim3\times10^{-3}\rm Mpc^{-3}$, and the luminosity averaged over the period $\tau$ is $\tau^{-1}E_\nu^2dN_\nu/dE_\nu\sim10^{38}\rm erg\,s^{-1}$. Both are well within the constraints by current IceCube data.

The neutrino luminosity of an individual SN event is, for confined wind case, $E_\nu L_{E_\nu}\approx E_\nu^2(dN_\nu/dE_\nu)/T\approx2.4\times10^{39} R_{w,15}^{5/8}\theta^{-8/7}(20E_\nu)\rm erg\,s^{-1}$. For a SN 10 Mpc away, the observed flux will be $2\times10^{-13}\rm erg\,cm^{-2}s^{-1}$, which might be detectable for future 10-Giga ton project Gen2 \citep{gen2}. Moreover, the accompanying gamma-ray flux from neutral pion decay is related to the neutrino flux as $E_\gamma^2\Phi_\gamma=2E_\nu^2\Phi_\nu(E_\gamma/2)\sim4\times10^{-13}\rm erg\,cm^{-2}s^{-1}$. The south CTA sensitivity is expected to be $\sim4\times10^{-14}\rm erg\,cm^{-2}s^{-1}$ for 50 hr exposure time at 3-10TeV range. The LHAASO sensitivity at 100 TeV for 1-yr exposure time is similar \citep{2016NPPP..279..166D}. Due to background free at $\ga100$~TeV for LHAASO, we can use scaling of $\propto 1/T$ to estimate the sensitivity for exposure time $T$, which is $\sim3\times10^{-13}(T/50\rm day)^{-1}\rm erg\,cm^{-2}s^{-1}$. So both CTA and LHAASO might be able to marginally detect a 10-Mpc event. The expected SN event rate within $10$~Mpc is $\sim3$ in 10 yrs. A follow-up observation by CTA or LHAASO for the core-collapse SNe within $\sim10$~Mpc is encouraging.

It should be noted that the accompanying high-energy gamma-rays may suffer pair-production absorption before escaping from the SN, and hence cannot be observed. We estimate the pair-production optical depth here. The main target photons for $\gamma\gamma$ interactions would be the thermal photons from the photosphere of the SN ejecta. The photon number density at radius $R$ is roughly $n_{ph}\sim L_{SN}/4\pi R^2(3kT)c$, with $L_{SN}$ the bolometric luminosity of the SN thermal radiation, and $T$ the temperature of the thermal radiation. For high-energy photons with $E_\gamma$, the threshold photon energy for pair-productions is $E_{th}\sim2(m_ec^2)^2/E_\gamma$. Thus the pair-production optical depth for $E_\gamma$ at $R$ is $\tau_{\gamma\gamma}\sim n_{ph}(\sigma_T/5)R(E_{th}/3kT)$ in the case of $3kT>E_{th}$. Typically $L_{SN}\sim10^{42}\rm erg\,s^{-1}$, and $T\sim10^4$~K, thus we have $\tau_{\gamma\gamma}\sim1.6R_{15}^{-1}(E_\gamma/10\rm TeV)^{-1}$, which implies that typically $E_\gamma\ga10$~TeV photons may escape partially. Moreover, the pair-production mean free path for high-energy photons propagating in the extragalactic background lights is $\lambda_{\gamma\gamma}\la20$~Mpc for $E_\gamma\ga10$~TeV \citep[see, e.g.,][]{Baret:2011zz}. In brief, the $10-100$~TeV photons accompanying the high-energy neutrinos from SNe II wind breakout events may be able to arrive the Earth avoiding significant absorption either in the sources or in propagation.

\section*{Notes added}
First, the very recent finding of delayed shock breakouts shows further evidences that most SNe II have dense circumstellar material \citep{Forster:2018mib}, supporting our assumption.

Second, very recently a neutrino event IceCube-170922A is claimed to be associated with a blazar TXS 0506+056, with a significance at $3\sigma$ level \citep{IceCube:2018dnn}. But more observations are needed to confirm this neutrino-blazar association for most extragalactic neutrinos.

\section*{Acknowledgments}

The author thanks Tian-Qi Huang and B.Theodore Zhang for helps in preparation, Kai Wang for discussion, and the several anonymous referees for helpful comments. This work is supported by the NSFC (No. 11773003) and the 973 Program of China (No. 2014CB845800).

\end{document}